\documentclass[a4paper]{wccm2020} \pdfoutput=1
\usepackage[numbers,square,sort&compress]{natbib}
\usepackage{xspace}
\usepackage[american]{babel}
\usepackage{xcolor,parskip,enumerate}
\usepackage{graphicx}
\usepackage{amsmath,amsfonts,amssymb,amsthm,bm}
\usepackage{tikzsymbols}        % \Sadey
\usepackage{dsfont}             % \mathds
\renewcommand{\mathbb}{\mathds} % \mathds --> \mathbb
\usepackage{hyperref}

\newtheorem{prop}{Proposition}
\theoremstyle{remark}
\newtheorem{remark}{Remark}

\newcommand   \Xset  {\mathbb{X}}

\newcommand   \RR    {\mathbb{R}}
\newcommand   \Rset  {\RR}

\newcommand   \Rplus {\RR_{+}}

\newcommand   \Ncal  {\mathcal{N}}
\newcommand   \Ucal  {\mathcal{U}}

\newcommand{\one}{{\mathbb{1}}}
\newcommand{\tr}{^{\mathsf{T}}}

\newcommand \Fs      {\ensuremath{F_{\mathrm{s}}}\xspace}

\newcommand \PropR   {(\mathrm{R})}
\newcommand \kstat  {k_{\mathrm{stat}}}
\renewcommand{\hat} {\widehat}
\newcommand \GP     {\mathrm{GP}}
\newcommand \mh     {\hat{m}}
\newcommand \xihat  {\hat{\xi}}

\newcommand \uxi    {\underline{\xi}}

\newcommand \metre  {\ensuremath{\mathrm{m}}}
\newcommand \Newton {\ensuremath{\mathrm{N}}}

\def\doctitle{%
  On the quantification of discretization uncertainty:
  comparison of two paradigms}

\def\dockw{%
  Quantification of uncertainty, %
  Discretization uncertainty, %
  Grid convergence index, %
  Gaussian processes, %
  Multi-fidelity}

\def\docabstract{%
  The use of simulation has spread to all areas of engineering and
  science, and the use of numerical models based on partial
  differential equations has thus multiplied. %
  The resolution of these models is generally based on the
  discretization of the space in which the solutions to the equations
  under consideration are sought. %
  The finite differences method or the finite elements method are two
  examples of such a discretization. %
  This discretization simplifies the solving but implies a form of
  uncertainty on the value of any quantity of interest. %
  To quantify this discretization uncertainty, the grid convergence
  index (GCI), based on the Richardson extrapolation technique, is now
  standard in the Verification and Validation (V\&V) literature. %
  But alternative approaches were also proposed in the statistical
  literature, such as Bayesian approaches with Gaussian process
  models. %
  The objective of this work is to compare on a standard test
  case from the literature (Timoshenko's beam) the well-established
  GCI-based approach to the---younger---Bayesian approach for the
  quantification of discretization uncertainty.}

\hypersetup{%
  pdftitle={\doctitle},%
  pdfauthor={Julien Bect, Souleymane Zio, Guillaume Perrin, Claire Cannamela, Emmanuel Vazquez},
  pdfsubject={\docabstract},
  pdfkeywords={\dockw},
  bookmarksnumbered=true,% Include section numbers in bookmarks
  hypertexnames=false,% Do not use guessable names for links
  citecolor=blue,% 
  colorlinks=true,%
  linkcolor=black,%
  breaklinks,%
  pdfborder=0 0 0}

\title{\doctitle}

\author{
  JULIEN BECT$^{1}$, %
  SOULEYMANE ZIO$^{1}$, %
  GUILLAUME PERRIN$^{2}$, %
  CLAIRE CANNAMELA$^{2}$ %
  AND EMMANUEL VAZQUEZ$^{1}$}

\heading{%
  Julien Bect, %
  Souleymane Zio, %
  Guillaume Perrin, %
  Claire Cannamela %
  and Emmanuel Vazquez}

\address{%
   $^{1}$ Université Paris-Saclay, CNRS, CentraleSupélec,\\
   Laboratoire des Signaux et Systèmes (L2S), 91190, Gif-sur-Yvette, France.
   \and
   $^{2}$ Commissariat à l'énergie atomique et aux énergies alternatives (CEA),\\
   CEA/DAM/DIF, 91297, Arpajon, France.
   \and
   $^{\star}$ Corresponding author.  E-mail: julien.bect@centralesupelec.fr}

\abstract{\docabstract}

\begin{document}

\section{INTRODUCTION}
\label{sec:intro}

Numerical models based on partial differential equations (PDE), or
integro-differential equations, are ubiquitous in engineering and
science, %
making it possible to understand or design systems for which physical
experiments would be expensive---sometimes impossible---to carry
out. %
Such models usually construct an approximate solution of the
underlying continuous equations, using discretization methods such as
finite differences or the finite elements method. %
The resulting discretization error introduces a form of uncertainty on
the exact but unknown value of any quantity of interest (QoI), %
which affects the predictions of the numerical model alongside other
sources of uncertainty such as parametric uncertainty or model
inadequacy (see, e.g., the typology proposed by~\cite{KO01}). %
The present article deals with the quantification of this
discretization uncertainty, %
which is an instance of the more general concept of \emph{numerical
  uncertainty} (see, e.g., chapters~7 and~8 of~\cite{RO2010book} and
references therein).

\begin{figure}
  \centering
  \includegraphics[width=13cm]{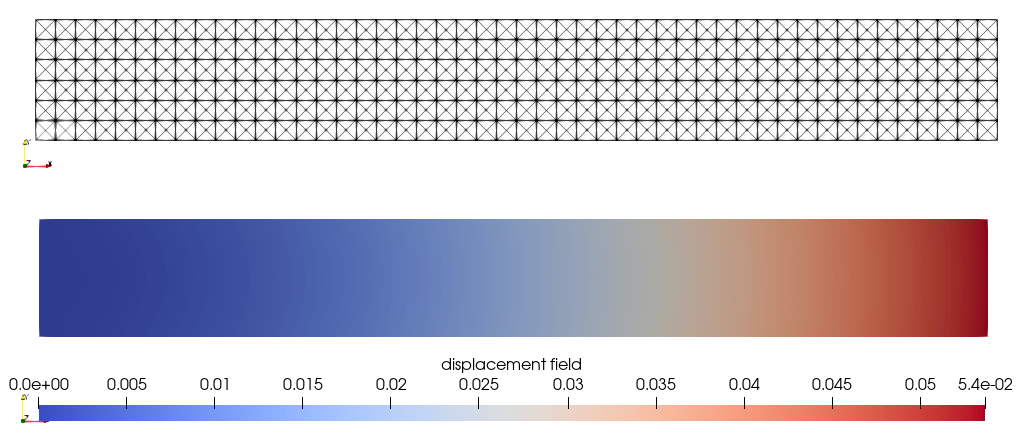}
  \vspace*{.5em}
  \caption{%
    Timoshenko's beam. %
    The isotropic rectangular beam is fixed at its left end, and a
    vertical pressure field is applied at the other end (see
    Section~\ref{subsec:timo} for details). %
    Top: discretization mesh with $L=48\metre$, $D=6\metre$ and
    $h = 1\metre$. %
    Bottom: displacement field for $\nu = 0.1$ and $P = 800$.}
  \label{fig:timo-beam}
\end{figure}

As an example, consider the ``Timoshenko beam'' problem (see
Figure~\ref{fig:timo-beam}), and assume that the QoI is the vertical
displacement of the beam, measured at a given location. %
The equations of linear elasticity describing this problem are
discretized in space using triangular mesh, the finesse of which is
controlled by parameter~$h$ . %
For a given value of the physical parameters of the problem (beam
dimensions, load, elasticity modulus, etc.), %
one run of the PDE solver returns an approximate value~$f(h)$ of the
QoI, where~$h$ denotes the particular value of the mesh parameter used
for this run. %
For convergent discretization schemes, the value~$f_0 \triangleq f(0)$ corresponds to
the value of the QoI for the exact solution of the equations, which
will never be perfectly known; it is usually possible, however, to run
the PDE solver for several values $h_1 > h_2 > \ldots > h_n$ of the
mesh parameter, in order to extrapolate $f_0$ from
$f(h_1), f(h_2),\ldots, f(h_n)$. %
The problem addressed in this article is the quantification of the
uncertainty on~$f_0$ given the results of such a grid refinement
study.

A first approach to this problem, now standard in the V\&V
(Verification and Validation) literature \cite{RO2010book}, %
uses the grid convergence index (GCI) originally proposed by
Roache~\cite{roache94} in the field of computational fluid dynamics
(CFD). %
It is based on a simple but clever reinterpretation of the Richardson
extrapolation technique, which has a long history in numerical
analysis, going back to the original work of
Richardson~\cite{richardson1911, richardson1927}. %
The key underlying assumption is that the discretization error behaves
as
\begin{equation}
  \label{eq:Rich-model}
  f(h) - f_0 \;=\; A h^p + o(h^p),
\end{equation}
where $A$ and~$p$ are two parameters that are usually considered
unknown (although a theoretical value of~$p$, known as the ``formal
order'' of convergence, is available in some situations). %
Roache proposed a method that uses~\eqref{eq:Rich-model}, not to
produce an extrapolated value~$\hat f_0$ as is usually done in
numerical analysis, but to construct an ``error band [\ldots] in which
the reader/user can have some practical level of confidence''
\cite[p.~407]{roache94}. %
The word ``confidence'', however, must be taken here---and later in
this article when referring to the GCI approach---in its casual,
non-statistical acceptation, %
and not as an indication of a ``well-founded probability statement''
\cite[p.~407 again]{roache94}.

Another approach was more recently introduced in the statistical
literature \cite{twy2014}, %
in the context of a general trend of research on the design and
analysis of computer experiments using Gaussian process (GP) models,
initiated at the end of the 80's by Sacks and co-authors
\cite{sacks89a, sacks89b, Currin91, Welch92}, %
and first applied to the analysis of computer experiments with
multiple levels of fidelity---also known as ``multi-fidelity''
computer experiments---by Kennedy and O'Hagan~\cite{KO01}. %
This approach relies on the Bayesian machinery: prior knowledge on the
function~$f: h \mapsto f(h)$ is encoded by a probability
distribution---a GP model for the sake of
tractability---, %
which gives birth, when combined with simulation results, to a
posterior distribution that quantifies the resulting uncertainty
on~$f$. %
In particular, this posterior distribution can be used to make
``well-founded probability statements'' (in a Bayesian sense) on the
unknown QoI~$f_0$.

The objective of this article is to present and compare these two
paradigms for the quantification of discretization uncertainty, which
have been developed in different scientific communities, %
and to assess the potential of the Bayesian approach to provide a
replacement for the well-established GCI-based approach, with better
probabilistic foundations. %
The article is organized as follows. %
Section~\ref{sec:two-paradigms} provides a short introduction to both
paradigms. %
Section~\ref{sec:cov-fun} provides theoretical results about several
classes of covariance functions that can be considered for the
Bayesian approach. %
Section~\ref{sec:num-exp} presents the result of our numerical results
on a standard test case from the literature---namely, Timoshenko's
beam. %
Finally, Section~\ref{sec:conclu} provides our conclusions and a
discussion of possible directions for future work.

\section{DISCRETIZATION UNCERTAINTY: TWO PARADIGMS}
\label{sec:two-paradigms}

This section summarizes the two main paradigms for the quantification
of discretization uncertainty. %
Other sources of numerical uncertainty (related, e.g., to the use of
iterative schemes to solve nonlinear equations) are assumed negligible
and will not be discussed in this article. %
In both cases, the numerical model under consideration will be assumed
to be deterministic. %
(The second approach can also deal with stochastic simulators; see,
e.g., the work of Stroh and co-authors
\cite{stroh2017firesafety,stroh2017AMCTM}.)

\subsection{Numerical analysis approach}
\label{sec:gci-approach}

In the scientific computing literature, the most commonly used method
for the quantification of discretization uncertainty is the GCI (Grid
Convergence Index) method, proposed by Roache
\cite{Roache1,Roachdisc,roache94} and reviewed, e.g., in Chapter~8 of
Roy and Oberkampf's book \citep{RO2010book}. %
It is based on a re-interpretation of Richardson's extrapolation
procedure \cite{richardson1911, richardson1927}, a well-established
idea in numerical analysis. %
More precisely, assume that the QoI satisfies
Equation~\eqref{eq:Rich-model} for some $A \in \Rset$ and $p > 0$. %
The values $f_0$, $A$ and~$p$ which appear in
Equation~\eqref{eq:Rich-model} can be estimated by evaluating~$f$ at the
different mesh sizes $h_1 < h_2 < h_3$ (often called fine, medium and
coarse) %
and then solving the system of nonlinear equations obtained by
neglecting higher order terms:
\begin{equation}
  f(h_k) = f_0 + A h_k^p,\quad 1 \le k \le 3.
\end{equation}
The solution can be written explicitly if the mesh sizes satisfy
$h_2 / h_1 = h_3 / h_2 = r$ for some~$r > 1$:
\begin{align*}
  \hat p & \;=\; \ln\left( \frac{f(h_3) - f(h_2)}{f(h_2) - f(h_1)}\right) / \ln(r),\\
  \hat A & \;=\; \frac{f(h_2) - f(h_1)}{h_1^{\hat p}\, \left( r^{\hat p} - 1 \right)},\\
  \hat f_0 & \;=\; f(h_1) + \frac{f(h_1) - f(h_2)}{r^{\hat p} - 1},
\end{align*}
yielding the approximation
\begin{equation}
  \label{equ:Richardson-approx}
  f(h) \;\approx\; \hat f(h) \;=\; \hat f_0 + \hat A\, h^{\hat p}.
\end{equation}
Until now what we have described is nothing more than Richardson's
extrapolation method. %
The GCI method takes the idea one step further, by considering a
confidence interval centered around the value of highest fidelity
(a.k.a.\ fine grid solution):
\begin{equation}
  \label{CI_gci}
  CI_{\text{GCI}} \;=\; \left[ f(h_{1}) - U;\; f(h_{1}) + U \right]
\end{equation}
for some $U > 0$.
Assuming that Equation~\eqref{equ:Richardson-approx} actually provides
an exact representation of~$f$---in other words, that there are no
higher-order terms---, it is then easy to see that the exact solution
belongs to the interval if, and only if,
\begin{equation}
  \label{CI_gci:U-formula}
  U =  U_{\text{GCI}} = \Fs\, \Bigl| \epsilon_{\text{GCI}} \Bigr|
  \qquad
  \text{for some $\Fs \ge 1$,}
\end{equation}
where $\epsilon_ {\text{GCI}}$ denotes the error made by using the
fine grid solution~$f(h_1)$ when~\eqref{equ:Richardson-approx} is
exact:
\begin{equation}
  \label{Runc}
  \epsilon_{\text{GCI}} \;=\;  \frac{\left| f(h_{1}) - f(h_{2}) \right|}{r^{\hat{p}}-1}.
\end{equation}

The constant $\Fs$ in Equation~\eqref{CI_gci:U-formula} is called the
``safety factor''. %
Roache \cite{roache94} recommends the use of~$\Fs = 3$ in general,
except when the value of~$p$ is known beforehand from numerical
analysis (in which case it is called the ``formal order of
convergence'') %
and it has been checked carefully that the solutions have been
computed in the ``asymptotic range'' where the
approximation~\eqref{equ:Richardson-approx} (with $\hat p = p$ known)
is accurate; in which case the value~$\Fs = 1.25$ is recommended.

\begin{figure}
  \hspace{-8mm}\includegraphics[width=17cm]{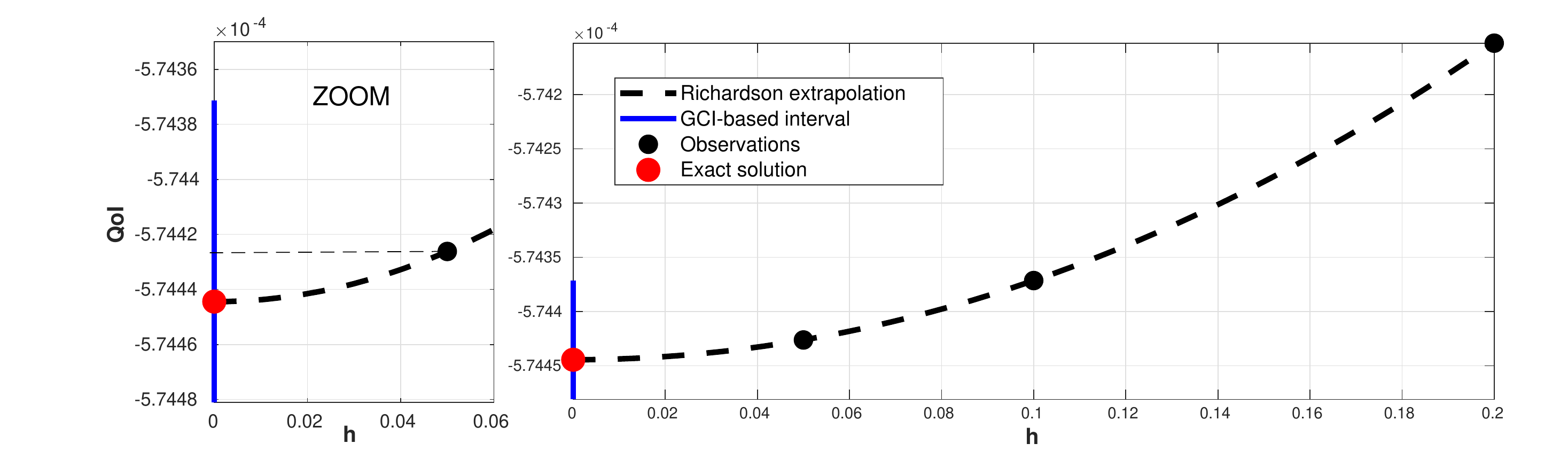}
  \caption{%
    Illustration of the GCI method on the Timoshenko beam problem,
    with $\Fs = 3$. %
    The QoI is the vertical displacement at~$x = 10\metre$. %
    Three grid sizes are used: $h_{1} = 1/20$, $h_{2}= 1/10$, and
    $h_{3} = 1/5$.}
  \label{fig:CIgci}
\end{figure}

Figure~\ref{fig:CIgci} illustrates the GCI method with an example
taken from the Timoshenko beam problem (see Section~\ref{subsec:timo}
for a full description). %
Observe in particular the interval is indeed centered around the fine
grid solution (and \emph{not} around the extrapolated solution): %
it is important to keep in mind that the GCI approach uses
Richardson's extrapolation technique to define a confidence
interval, but not to actually extrapolate to a more accurate solution.

\subsection{Probabilistic (Bayesian) approach}
\label{sec:Bayesian-approach}

In 2014, Tuo and co-authors \cite{twy2014} introduced an approach to deal with\footnote{%
  To be precise, \cite{twy2014} considers a QoI that depends on a
  vector~$x$ ranging in a set~$\Xset$ corresponding to the input space
  of a numerical simulator. For a given~$x\in\Xset$, the objective is to
  infer~$f(x, 0)$ given simulation results
  $f(x_1, h_1), \ldots, f(x_n, h_n)$. Here, we focus on a more specific
  problem, that is, \emph{grid refinement studies}, where $x$ is fixed
  and $h$ varies.}
discretization uncertainty using a concept of Bayesian
multifidelity~\citep{KO01, pichy}. %
Under this approach, the QoI $f_0$ is assumed to be a sample value of a
random variable~$\xi_0$, which models uncertainty about the unknown
value~$f_0$. %
The \emph{prior} distribution of this random variable---i.e., its
distribution before any actual evaluation of the numerical model is
made---encodes prior beliefs about the plausible values of~$f_0$.

In the domain of design and analysis of computer experiments, it is
customary to use Gaussian prior distributions for the outputs of
(deterministic) numerical simulators, since Gaussian distributions are
conjugate priors, which conveniently lead to Gaussian posterior
distributions. %
Following~\cite{twy2014}, we model the QoI~$f(h)$ at mesh size~$h>0$
using a random process~$\xi$, such that,
\begin{equation}
  \label{eq:bayesian-model}
  \forall h \in \Rplus,\quad
  \xi(h) = \xi_0 + \varepsilon(h),
\end{equation}
where $\xi_0 \sim \Ncal(m_0, \sigma_0^2)$ for some
hyper-parameters~$m_0 \in \Rset$ and~$\sigma^2 > 0$, %
and $\varepsilon$ denotes a zero-mean GP, independent of
$\xi_0$, which is assumed to converge to zero in the mean-square sense
when $h$ goes to zero:
\begin{equation}
  \label{eq:mf-property}
  \mathop{\rm var} \left( \varepsilon(h) \right) \;\xrightarrow[h\to 0]{}\; 0.
\end{equation}
The GP $\varepsilon$ corresponds to the error of
discretization---in other words, $\varepsilon$ models the \emph{loss of
  fidelity} as $h$ increases. % 
The  distribution of $\varepsilon$ will be denoted by
$\GP(0, k_{\varepsilon})$, where $k_{\varepsilon}$ stands for the
covariance function of~$\varepsilon$, which is such that
$\lim_{h,h^{\prime}\to 0} k_{\varepsilon}(h,h^{\prime}) = 0$, due
to~(\ref{eq:mf-property}). %
Note that, conditional on $\xi_0$, $\xi$~is a non-stationary GP
with mean function~$\xi_0 \one_{\Rplus}$ and covariance
function~$k_{\varepsilon}$.

In practice, it is convenient to assume an improper uniform
distribution~$\Ucal_{\Rset}$ for~$\xi_0$, which may be thought as
taking the limit $\sigma_0^2 \to \infty$ (for any
fixed~$m_0 \in \Rset$). %
In this case, the posterior distribution of~$\xi$ is given by the
equations of \emph{ordinary kriging}, which are recalled in the
following.

\begin{prop}[Ordinary kriging]
  \label{prop:gauss-proc-priors}
  Let $\Xset$ denote a set and $k$ a covariance function on~$\Xset$. %
  Let $\xi$ denote an (improper) GP on~$\Xset$, such
  that $\xi \mid m \;\sim\; \GP\left( m,\, k \right)$ and
  $m \sim \Ucal_\Rset$. %
  Let $n \ge 1$ and $h_1,\, \ldots,\, h_n \in \Xset$. %
  Then, for all $h \in \Xset$,
  \begin{equation*}
    \xi(h) \mid \xi(h_1), \ldots, \xi(h_n)\;\sim\; 
    \Ncal\left( \xihat_n(h),\, s_n^2(h) \right)\,, \vspace{-5pt}
  \end{equation*}
  with \vspace{-5pt}
  \begin{align}
    \label{eq:pred}
    \xihat_n (h) & \;=\;  \mh_n + k_n(h)\tr K_n^{-1}(\uxi_n - \mh_n \one_n)\,,\\
    \label{eq:variance}
    s_n^2(h) & \;=\;
      k(h,h) \,-\, k_n(h)\tr K_n^{-1} k_n(h) +
      \frac{(1 - k_n(h)\tr K_n^{-1}\one_n)^2}{\one_n\tr K_n^{-1} \one_n},
  \end{align}
  where $\uxi_n=(\xi(h_1), \ldots, \xi(h_n))\tr$ is the vector of
  observations, %
  $k_n(h)$ the correlation vector between $\xi(h)$ and~$\uxi_n$, %
  $K_n$ the covariance matrix of~$\uxi_n$, %
  $\one_n = (1,\ldots, 1)\tr$, %
  and
  $\mh_n = \one_n\tr K_n^{-1} \uxi_n \,/\, \one_n\tr K_n^{-1} \one_n$
  the weighted least squares estimate of~$m$.
\end{prop}

\bigbreak

Proposition~\ref{prop:gauss-proc-priors} provides us with a method to build
confidence interval about $f_0$ from simulations at mesh sizes
$h_1, \ldots, h_n$. %
% The choice of the mesh sizes, which is called the Design of
% Experiments (DoE) is discussed in Section~\ref{sec:num-exp}. %
The procedure to compute confidence (the term \emph{credibility} would
be preferred under a Bayesian terminology) intervals consists of the
following steps:
\begin{enumerate}
\item Given simulation results $f(h_1), \ldots, f(h_n)$, and the
  choice of a parameterized covariance function $k$, which will be
  discussed in Section~\ref{sec:cov-fun}, estimate the parameters of
  $k$ using a maximum likelihood approach (see
  Section~\ref{sec:num-exp}).\footnote{%
    The proposed procedure does not take into account the uncertainty
    resulting from the estimation of the parameters of the covariance.
    This could be investigated in future work.}

\item Using the covariance function $k$ estimated at the previous step
  and the corresponding GP model~$\xi$ for~$f$, compute the Gaussian
  posterior distribution of the QoI $\xi_0 = \xi(0)$. %
  This gives a posterior mean value $\xihat_n(0)$, which corresponds
  to~$\mh_n$ since $k$~goes to zero at the origin, and a posterior
  variance $s_n^2(0)$. %
  (Of course, the user can in fact obtain the posterior mean and the
  posterior variance at any $h$.)
\item Using $\xihat_n(0)$ and $s_n^2(0)$, derive a credibility
  interval at level $\alpha \in (0,1)$ under the form
  \begin{equation*}
    CI_{\text{MF}} \;=\; \biggl[\xihat_n(0) + s_n(0)\Phi^{-1}\Bigl(\frac{1-\alpha}{2}\Bigr);\,
    \xihat_n(0) + s_n(0)\Phi^{-1}\Bigl(\frac{1+\alpha}{2}\Bigr)\biggr]\,,
  \end{equation*}
  where $\Phi^{-1}$ stands for the quantile function of the normal
  distribution.
  % Standard values for $\Phi^{-1}\Bigl(\frac{1-\alpha}{2}\Bigr) = -
  % \Phi^{-1}\Bigl(\frac{1+\alpha}{2}\Bigr)$ are recalled in
  % Table~\ref{tab:quantiles}. (The notation ``MF'' stands for
  % multi-fidelity.)
\end{enumerate}

\begin{figure}
  \centering
  \includegraphics[width=0.9\textwidth]{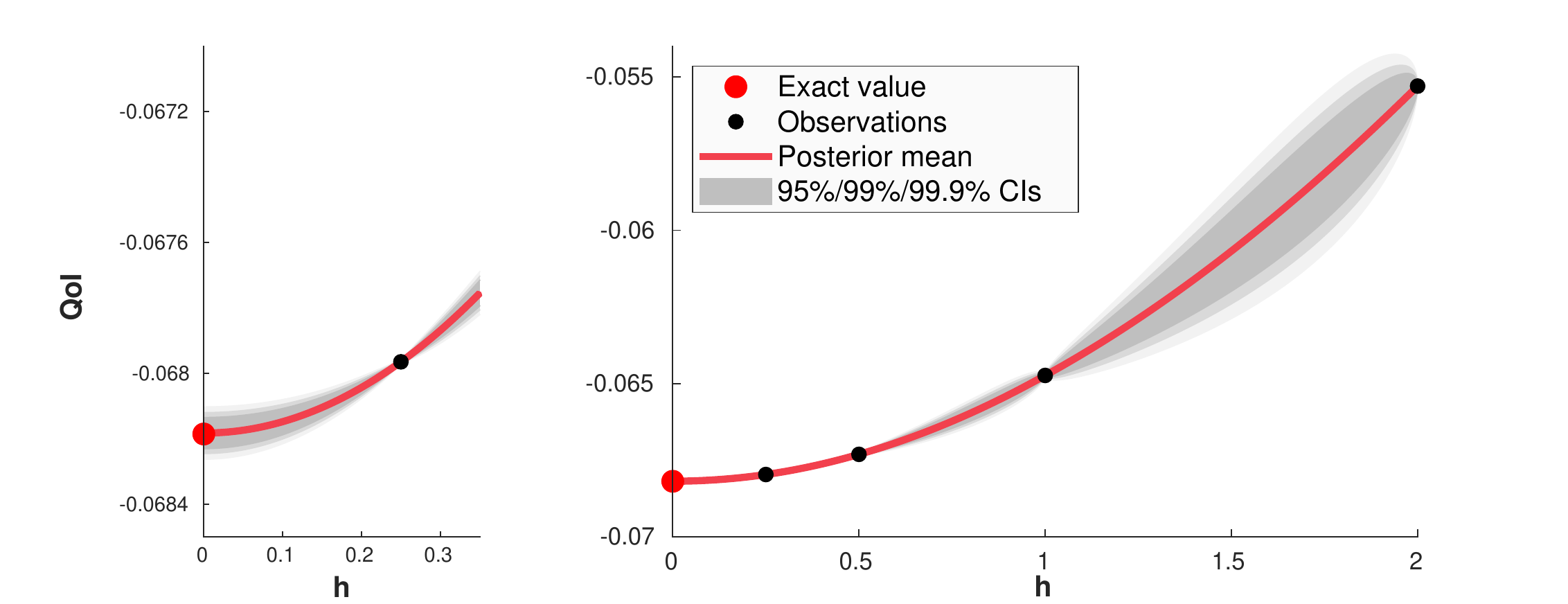}
  \caption{%
    Illustration of the Bayesian approach. %
    Left: global view.  Right: zoom at~$h = 0$. %
    Four observations, at $h = 2, 1, 0.5, 0.25$. %
    GP model: TWY2 covariance function (see Section~\ref{sec:cov-fun})
    with $\sigma = 5\, 10^{-3}$, $L = 4$ and a Matérn-$1/2$
    (exponential) stationary covariance function with range
    parameter~$\rho = 200$.}  
  \label{fig:illustr-gp}
\end{figure}

The procedure is illustrated on Figure~\ref{fig:illustr-gp}. %
Observe that, in contrast with the GCI approach, the confidence (or
credibility) intervals produced by the Bayesian approach are not, in
general, centered around the highest-fidelity value. %
They are centered instead around the extrapolated value, that is, the
posterior mean at~$h = 0$. %
In the special case of the Brownian-like covariance function
of~\cite{twy2014} (see below), however, the extrapolated value
coincides with the observation of highest fidelity.

\section{COVARIANCE FUNCTIONS}
\label{sec:cov-fun}

In this section, we tackle from a theoretical point of view the
question of choosing a suitable covariance function for the
GP model in the Bayesian approach of
Section~\ref{sec:Bayesian-approach}, %
when prior evidence indicates that the unknown function~$f$
obeys~\eqref{eq:Rich-model} as in the GCI approach. %
More precisely, assuming that~$\varepsilon$ is a zero-mean
GP with continuous sample paths\footnote{%
  Let us recall a classical sufficient condition for the sample path
  continuity of a zero-mean GP on $\Xset \subset \Rset^d$ \citep[see,
  e.g.,][Theorem~3.4.1]{Adler81}: if there exist $C > 0$ and
  $\eta > 0$ such that
  $k(x,x) + k(y,y) - 2 k(x,y) \le \frac{C}{\left| \log \lVert x - y
      \rVert \right|^{1 + \eta}}$ for all $x,y \in \Xset$, %
  then there exists a version of~$\xi$ with continuous sample paths.
  % This is a very weak condition, which is satisfied by all commonly
  % used continuous covariance functions on~$\Rset^d$ (e.g.,
  % geometrically anisotropic or tensor-product Matérn covariance
  % functions).
} on~$\left[ 0, +\infty \right)$, %
we will provide conditions on the covariance function~$k$ under which the property
\begin{equation*}
  \PropR:\qquad
  \exists A \neq 0, \;
  \exists p > 0,\quad
  \varepsilon(h) = A h^p + o(h^p),
\end{equation*}
holds almost surely %
(where~$A$ and~$p$ are allowed to depend on the particular sample path
that is considered). %
Proofs are omitted for the sake of brevity, and will be provided in a
forthcoming publication. %
Our first result deals with the ``Brownian-like'' model that is
recommended in~\cite{twy2014}, and shows that this model is not, in
fact, a suitable prior for this type of problem.

\begin{prop} \label{prop-TWY1}
  Assume that $k$ is of the form
  \begin{equation*}
    \tag{TWY1}
    k(h, h') =
    \sigma^2\, \min (h, h') ^L,
  \end{equation*}
  with $\sigma^2$ and~$L$ some positive parameters. %
  Then, almost surely, $\PropR$ does not hold.
\end{prop}

Our second result deals with the second model that is considered---but
not advocated---in~\cite{twy2014}.

\begin{prop} \label{prop-TWY2}
  Assume that~$k$ is of the form
  \begin{equation*}
    \tag{TWY2}
    k(h, h') =
    \sigma^2\; (h h')^{L/2}\; c(h - h'),
  \end{equation*}
  where $\sigma^2$ and~$L$ are positive parameters, and $c$ is the
  stationary correlation function of a GP with continuous sample
  paths. %
  Then $\PropR$ holds almost surely with $p = \frac{L}{2}$.
\end{prop}

This model is thus suitable for the problem under consideration,
for any value of~$p$.
% The value of~$p$ must be known beforehand, since it appear as a
% parameter in the covariance function, or estimated from the data (see
% Section~\ref{sec:num-exp:estim-L} for further discussion).
Note that, for a given value of~$p$, significant modeling flexibility
remains, through the choice of the stationary correlation~$c$. %
For instance, the squared exponential (a.k.a.\ Gaussian) correlation
can be used as in~\cite{twy2014}, but rougher correlation functions,
such as the Matérn family of correlation functions~\citep{stein99},
can be considered as well.

Our last result deals with covariance functions of the form
\begin{equation}
  \label{eq:stat-case-k}
  k(h, h') = \sigma^2 \left[ 1 + c(h-h') - c(h) - c(h') \right],
\end{equation}
where $c$ is a stationary correlation function. %
This is the covariance function of~$\tilde \xi - \tilde \xi (0)$, %
where $\tilde \xi$ is a stationary GP with covariance
function~$\kstat(h, h') = \sigma^2\, c(h - h')$.

\begin{prop} \label{prop-STZ} Assume that $k$ is of the
  form~\eqref{eq:stat-case-k}, where $\sigma^2$ is a positive
  parameter and $c$ is the stationary correlation function of a GP
  with continuous sample paths.
  \begin{enumerate}[i) ]
  \item If property~$\PropR$ holds almost surely, then $p \le 1$ almost
    surely.
  \item If $c$ is the covariance function of a GP with differentiable
    sample paths, then there exists a version of~$\varepsilon$ such that
    property~$\PropR$ holds almost surely with~$p = 1$.
  \end{enumerate}
\end{prop}

Such covariance functions are thus, in principle, only suitable
for~$p \le 1$. %
(We even conjecture that \ref{prop-STZ}.i~actually holds with
``$p = 1$'' instead of ``$p \le 1$'').

\section{NUMERICAL EXPERIMENTS}
\label{sec:num-exp}

\subsection{Test case}
\label{subsec:timo}

We consider the quasi-static deformation of an
isotropic rectangular beam.
$\Omega=[0,L]\times[-D/2,D/2]$ is the domain characterizing the
initial position of the beam, $\Gamma_\text{left}$ and
$\Gamma_\text{right}$ are the left and right sides of the beam, and
$\Gamma_\text{other}$ is the union of the two other sides of the
beam. %
The beam is supposed to be fixed at one end, and a specific vertical
pressure field, denoted by $p$, is applied at the other end. %
Under linear elasticity, and neglecting the gravity forces
and the atmospheric pressure, it can be shown \cite{timo} that the
displacement field in each point of the beam, written
$\mathbf{u}=(u_1,u_2)$, can be modeled by the solution of the
following system of equations:
\begin{equation}\label{govT} \small
  \left\{
    \begin{aligned}
      & \mathrm{\nabla}\,{\cdot}{\sigma}(\mathbf{u}) = \mathbf{0}\\
      & \mathbf{u} = \mathbf{0}\\
      & {\sigma}(\mathbf{u})\cdot \mathbf{e}_2 = \mathbf{0}\\
      & {\sigma}(\mathbf{u})\cdot \mathbf{e}_1 = (0, -p)
    \end{aligned}
  \right.  \ \
  \begin{aligned}
    & \mbox{in}\: \Omega,\\
    & \mbox{on}\: \Gamma_{\text{left}},\\
    & \mbox{on}\: \Gamma_{\text{other}},\\
    & \mbox{on}\: \Gamma_{\text{right}},
  \end{aligned}
\end{equation}
where
$\sigma(\mathbf{u})=\lambda
\text{Trace}(\varepsilon(\mathbf{u}))\mathbf{I}+2\mu
\varepsilon(\mathbf{u})$,
$\varepsilon(\mathbf{u})=\frac{1}{2}\big(\nabla \mathbf{u} + \nabla
\mathbf{u}^\top \big)$, and $\lambda,\mu$ are two parameters
characterizing the material properties of the beam. %
If the pressure field~$p$ applied at the free end of the beam
($x_1 = L$) is given by %
$p =\frac{P}{2I}\big( \frac{D^{2}}{4} - x_{2}^{2} \big)$ %
with $I = \frac{1}{12}\, D^3$ the moment of inertia of the beam, there exists an explicit
solution for $\mathbf{u}$:
\begin{eqnarray}\label{uxeact} \small
  u_{{1}}(x_1,x_2) &\;=\;
  & \frac{P\,x_{2}}{6\,E^{*}\,I}\, \left[
    (6\,L - 3\,x_{1})x_{1} +(2+\nu^{*})x_{2}^{2} - \frac{3D^{2}}{2}(1+\nu^{*})
    \right],\\
  u_{{2}}(x_1,x_2) &\;=\;
  & \frac{P}{6\,E^*\,I}\, \left[
    3\,\nu^{*} \,x_{2}^{2}(L - x_{1}) +(4+5\nu^{*})\frac{D^{2}x_{1}}{4} +(3\,L -x_{1} )x_{1}^{2}
    \right],
\end{eqnarray}
with $\lambda = \frac{E\nu}{(1+\nu)(1-2\nu)}$,
$\mu = \frac{E}{2(1+\nu)}$, $E^{*}=\frac{E}{1-\nu}$ and
$\nu^{*}=\frac{\nu}{1-\nu}$. %
In the numerical experiments, the values of $E$, $L$ and~$P$ will be fixed to
$3\times10^{7}\, \Newton/\metre^{2}$, $48\metre$ and~$1000\Newton$ respectively,
whereas the values of~$D$ and $\nu$ will vary.
Finally, we will consider as QoIs the vertical displacement at four
different positions:
\begin{equation*}
  f_j = u_2 \left( x_{p_j}, 0 \right), \qquad 1 \le j \le 4,
\end{equation*}
with $x_{p_1} = 10 \metre$, $x_{p_2} = 20 \metre$,
$x_{p_3} = 30 \metre$ and $x_{p_4} = 48 \metre$.

\begin{remark}
  Augarde and Deeks \cite{augarde2008use} discuss alternative
  formulations of the beam problem with more realistic boundary
  conditions (but no analytical solutions).
\end{remark}

\begin{figure}
  \centering \includegraphics[scale=0.55]{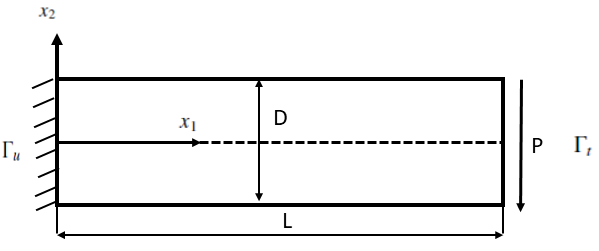}
  \caption{Timoshenko Beam}
  \label{fig:beam:representation}
\end{figure}

\subsection{Experimental setup}
\label{subsec:exp-setup}

We consider six possible values for the height: %
$D \in \{ 2\metre, 4\metre, \ldots, 12\metre \}$ %
and nine values for the Poisson ratio: %
$\nu \in \{0, 0.10, 0.15, 0.20, 0.25, 0.30, 0.35, 0.40 ,0.45\}$, %
thereby creating 54~instances of the Timoshenko beam problem. %
The PDE is solved in FEniCS \cite{Fenics1}, using a finite element method
with a regular triangular mesh (see Figure~\ref{fig:timo-beam}).

Two methods for constructing confidence intervals about the exact
value of QoIs are compared: %
the GCI approach (see Section~\ref{sec:gci-approach}) with
$\Fs = 3$ on the one hand, and the Bayesian approach (see
Section~\ref{sec:Bayesian-approach}) with~$\alpha = 99.9\%$ on the
other hand. %
These large values of~$\Fs$ and~$\alpha$ have been chosen in order to
construct conservative intervals, which are thus expected to contain
the true value in most (if not all) instances. %
We use a 3-point DoE with $r = 2$ for the GCI method:
$h^{\text{GCI}}_j = 2^{j-3} \cdot h^{\text{GCI}}_3$, $1 \le j \le 3$,
with $h^{\text{GCI}}_3 = \frac{L}{216} = \frac{2}{9}\metre$, %
and a $16$-point DoE for the GP method:
$h^{\text{GP}}_j = h^{\text{GP}}_n / (n+1-j)$, $1 \le j \le n$, with
$n=16$ and $h^{\text{GP}}_{16} = \frac{L}{24} = 2\metre$. %
These DoEs have been chosen in order to have similar computation times
for both methods.

For the Bayesian approach, three classes of covariance functions are
considered: the two classes of covariance functions proposed
by~\cite{twy2014} and described in
Propositions~\ref{prop-TWY1}--\ref{prop-TWY2}, denoted as TWY1 and
TWY2 respectively (for ``Tuo, Wu and Yu''), and the one described in
Proposition~\ref{prop-STZ}, denoted by STZ (for ``STationary minus the
value at Zero''). %
For the TWY1 and TWY2 classes, the parameter~$L$ is either set
to~$L=1$ or~$L=2$, or estimated along with the others by restricted
maximum likelihood. %
For the TWY2 and STZ classes, a stationary correlation function must
be specified: we consider as possible choices the Matérn covariance
function with regularity~$\nu = \frac{1}{2}$ (a.k.a. exponential
covariance function), $\nu = \frac{3}{2}$ or~$\nu = \frac{5}{2}$, the
Matérn covariance function with estimated regularity, and the Gaussian
(a.k.a. squared exponential) covariance function. %
All in all, this gives us a total of 23~covariance models to be
compared (3 for the TWY1 class, 15 for the TWY2 class and 5 for the
STZ class). %
GP modeling computations are carried out using the STK toolbox
\cite{stk}.

\subsection{Results}
\label{subsec:results}

\begin{figure} \centering
  \includegraphics[width=0.53\textwidth]{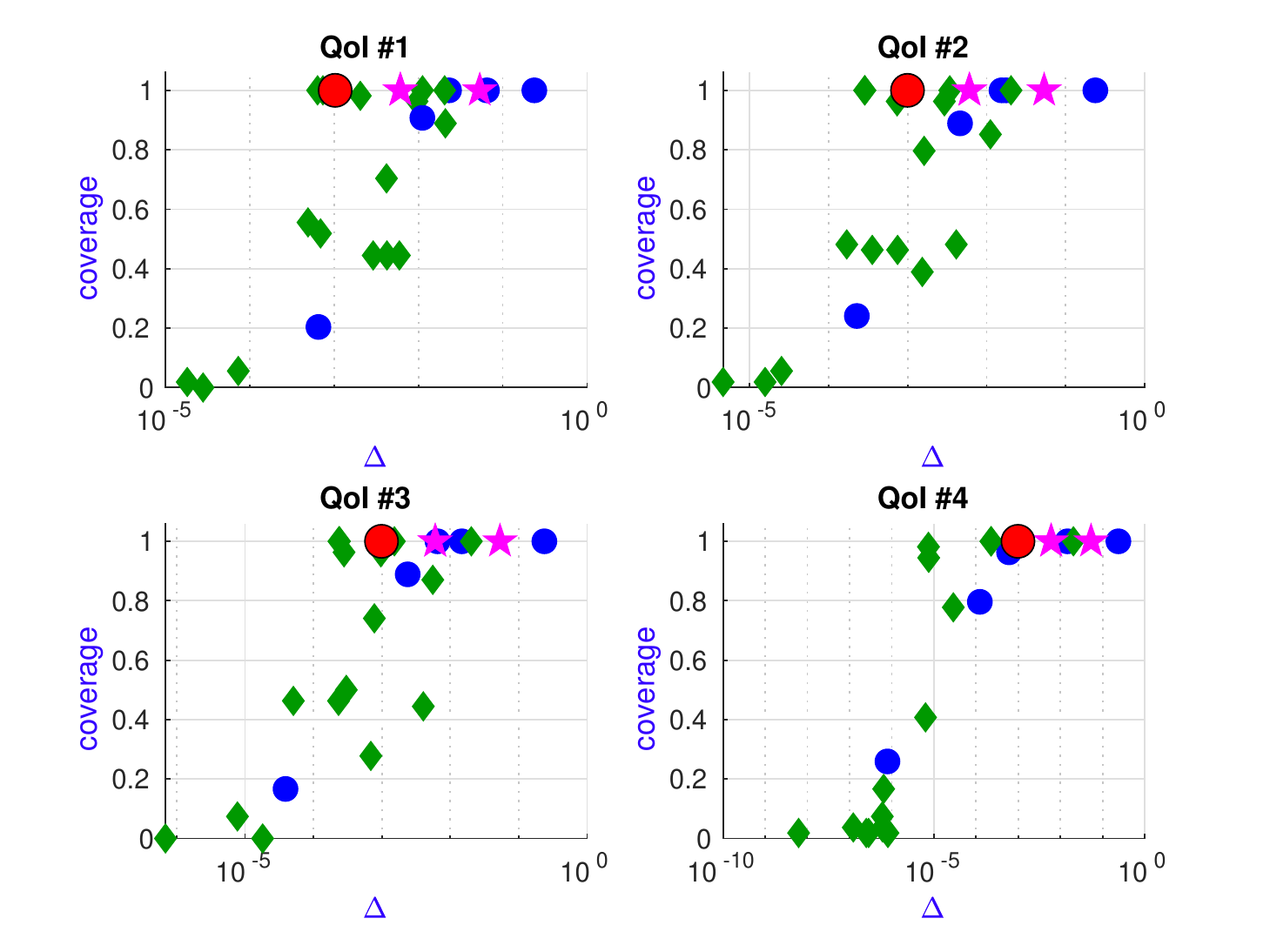}
  \caption{%
    Empirical coverage and average interval half-width for the 23+1=24
    methods. %
    Red disks: GCI-based intervals. %
    Pink stars: TWY1. %
    Green diamonds: TWY2. %
    Blue disks: STZ.}
  \label{fig:cov-hw}
\end{figure}

Figure~\ref{fig:cov-hw} presents the average performance of all the
23+1=24 methods on the 54 instances of the problem. %
Two performance metrics are considered: %
the coverage of the interval, which is the proportion of instance
where the interval contains the true value, %
and the average interval half-width (IWH), denoted by~$\Delta$. %
As expected, the GCI interval is conservative---its empirical coverage
is actually equal to 100\% in this experiment, for all four QoIs. %
The results are much more contrasted for the Bayesian approach,
however: depending on the covariance function, the interval is either
overconfident (small but with a low coverage), or over-conservative
(vary large interval), or---and this is the interesting
case---simultaneously smaller than the GCI interval and with a good
coverage.

A closer look at the results allows to identify some promising classes
of covariance functions for the Bayesian approach. %
Table~\ref{tab:metrics} provides a more detailed view of the
performances of the 23 covariance models, focusing on the most
promising ones---defined as those which obtained both a reasonably
high coverage (80\% or more) and a reasonably low average IWH
($\Delta \le 3 \Delta_{\mathrm{GCI}}$). %
A first, striking observation is that, in this study, only one class
of covariance models manages to deliver intervals that are
simultaneously smaller than the GCI interval and with a good
coverage: %
the TWY2 class, and more precisely the TWY2 class with a
weakly-regular stationary correlation function (Matérn
with~$\nu = \frac{1}{2}$ or~$\nu = \frac{3}{2}$).

Concerning the decay parameter~$L$, the best performances are obtained
when it is fixed a priori to~$L = 4$, or estimated. %
This is consistent with Proposition~\ref{prop-TWY2}, since it
corresponds to a quadratic convergence ($p = \frac{L}{2} = 2$), which
is the actual convergence rate in this problem. %
Note that, with the Matérn-3/2 covariance function, the results when
$L$~is estimated are not as good as when it is fixed to the
true value, which is not surprising per se, but suggests that
there might be room for improvement in the parameter selection
procedure. %
The fact that covariance functions from the TWY1 and STZ classes do not lead to
satisfactory intervals is also consistent with the theoretical results
of Section~\ref{sec:cov-fun}. %
Indeed, from a sample path point of view, neither of these classes
provides a suitable prior for a quadratic convergence at~$h = 0$
(see Propositions~\ref{prop-TWY1} and~\ref{prop-STZ}, respectively).

\newcommand\bad{\Sadey}
\newcommand\MM[1]{$\mathcal{M}_{#1}$}

\begin{table}\small
  \centering
  \begin{tabular}[h]{|l|c|c|c|c|}
    \hline
    covariance
    & QoI \#1
    & QoI \#2
    & QoI \#3
    & QoI \#4
    \\ \hline
    TWY2 (\MM{1/2}, $L = \hat L$)
    & 1.000, \textcolor{blue}{0.716}
    & 1.000, \textcolor{blue}{0.287}
    & 1.000, \textcolor{blue}{0.239}
    & 1.000, \textcolor{blue}{0.228}
    \\
    TWY2 (\MM{1/2}, $L=4$)
    & 1.000, \textcolor{blue}{0.619}
    & 1.000, \textcolor{blue}{0.289}
    & 1.000, \textcolor{blue}{0.243}
    & 1.000, \textcolor{blue}{0.232}
    \\    
    TWY2 (\MM{3/2}, $L = \hat L$)
    & \bad
    & 0.963, \textcolor{blue}{2.931}
    & 0.963, \textcolor{blue}{0.980}
    & 0.944, \textcolor{blue}{0.008}
    \\
    TWY2 (\MM{3/2}, $L=4$)
    & 0.981, \textcolor{blue}{1.989}
    & 0.963, \textcolor{blue}{0.739}
    & 0.963, \textcolor{blue}{0.284}
    & 0.981, \textcolor{blue}{0.008}
    \\
    TWY2 (\MM{3/2}, $L=2$)
    & \bad
    & \bad
    & 1.000, \textcolor{blue}{1.543}
    & 1.000, \textcolor{blue}{0.637}
    \\
    STZ (\MM{5/2})
    & \bad
    & \bad
    & \bad
    & 0.963, \textcolor{blue}{0.614}
    \\
    STZ (\MM{\hat\nu})
    & \bad
    & \bad
    & 0.889, \textcolor{blue}{2.414}
    & \bad
    \\
    All the other cases
    & \bad
    & \bad
    & \bad
    & \bad
    \\ \hline
  \end{tabular}
  \medbreak  
  \caption{%
    Performance metrics for the GP-based variants. %
    For each covariance function and each QoI: the empirical coverage
    is given first, and then (in blue) the ratio
    $\Delta / \Delta_{\mathrm{GCI}}$ of the IHW of the GP-based method
    by the IHW of the GCI-based method. %
    \MM{\nu} denotes the Matérn covariance with regularity~$\nu$. %
    The sad face~\bad indicates that either the coverage is below $80\%$
    or $\Delta > 3 \Delta_{\mathrm{GCI}}$.}  \label{tab:metrics}
\end{table}

\section{CONCLUSIONS}
\label{sec:conclu}

The Bayesian (GP-based) approach to the quantification of
discretization uncertainty emerges from this study as a promising
alternative to the GCI approach, %
with the potential to provide ``well-founded probability statements''
and, ultimately, better confidence intervals (i.e., shorter intervals
that still have a satisfactory coverage).

At the present time, however, the Bayesian approach lacks the maturity
and robustness of the GCI approach: which covariance model to use, and
how to robustly select the hyper-parameters (e.g, the decay
parameter~$L$ in the TWY2 model), are important questions that deserve
further attention. %
% Concerning the former, both (elementary) theoretical considerations
% and numerical results hint at the TWY2 class being a sensible
% choice. %
% Concerning the latter, fully Bayesian methods with objective priors
% and cross-validation techniques could be interesting directions for
% future work.
In a different direction, the construction of (possibly sequential)
DoEs, both for grid-refinement studies as considered in this article,
and for the more general case of parametric studies, is also an
interesting direction for future work.

\bibliographystyle{plainnat}
\bibliography{wccm2020-gcigp}

\begin{thebibliography}{21}
\providecommand{\natexlab}[1]{#1}
\providecommand{\url}[1]{\texttt{#1}}
\expandafter\ifx\csname urlstyle\endcsname\relax
  \providecommand{\doi}[1]{doi: #1}\else
  \providecommand{\doi}{doi: \begingroup \urlstyle{rm}\Url}\fi

\bibitem[Adler(1981)]{Adler81}
R.~J. Adler.
\newblock \emph{The Geometry of Random Fields}.
\newblock Wiley, New York, 1981.

\bibitem[Aln{\ae}s et~al.(2015)Aln{\ae}s, Blechta, Hake, Johansson, Kehlet,
  Logg, Richardson, Ring, Rognes, and Wells]{Fenics1}
M.~S. Aln{\ae}s, J.~Blechta, J.~Hake, A.~Johansson, B.~Kehlet, A.~Logg,
  C.~Richardson, J.~Ring, M.~E. Rognes, and G.~N. Wells.
\newblock The fenics project version 1.5.
\newblock \emph{Archive of Numerical Software}, 3\penalty0 (100), 2015.

\bibitem[Augarde and Deeks(2008)]{augarde2008use}
C.~E. Augarde and A.~J. Deeks.
\newblock The use of timoshenko's exact solution for a cantilever beam in
  adaptive analysis.
\newblock \emph{Finite elements in analysis and design}, 44\penalty0
  (9--10):\penalty0 595--601, 2008.

\bibitem[Bect et~al.(2019)Bect, Vazquez, et~al.]{stk}
J.~Bect, E.~Vazquez, et~al.
\newblock {STK}: a {S}mall ({M}atlab/{O}ctave) {T}oolbox for {K}riging.
  {R}elease 2.6.1, 2019.
\newblock URL \url{http://kriging.sourceforge.net}.

\bibitem[Currin et~al.(1991)Currin, Mitchell, Morris, and Ylvisaker]{Currin91}
C.~Currin, T.~J. Mitchell, M.~Morris, and D.~Ylvisaker.
\newblock {B}ayesian prediction of deterministic functions, with applications
  to the design and analysis of computer experiments.
\newblock \emph{Journal of the American Statistical Association}, 86\penalty0
  (416):\penalty0 953--963, 1991.

\bibitem[Kennedy and O'Hagan(2001)]{KO01}
M.C. Kennedy and A.~O'Hagan.
\newblock {B}ayesian calibration of computer models.
\newblock \emph{Journal of the Royal Statistical Society. Series B Statistical
  Methodology}, 63\penalty0 (3):\penalty0 425--464, 2001.

\bibitem[Oberkampf and Roy(2010)]{RO2010book}
W.~L. Oberkampf and C.~J. Roy.
\newblock \emph{Verification and validation in scientific computing}.
\newblock Cambridge University Press, 2010.

\bibitem[Picheny and Ginsbourger(2013)]{pichy}
V.~Picheny and D.~Ginsbourger.
\newblock A nonstationary space-time {G}aussian process model for partially
  converged simulations.
\newblock \emph{SIAM/ASA Journal on Uncertainty Quantification}, 1\penalty0
  (1):\penalty0 57--78, 2013.

\bibitem[Richardson(1911)]{richardson1911}
L.~F. Richardson.
\newblock The approximate arithmetical solution by finite differences of
  physical problems involving differential equations, with an application to
  the stresses in a masonry dam.
\newblock \emph{Philosophical Transactions of the Royal Society of London.
  Series A, Containing Papers of a Mathematical or Physical Character},
  210\penalty0 (459-470):\penalty0 307--357, 1911.

\bibitem[Richardson and Gaunt(1927)]{richardson1927}
L.~F. Richardson and J.~A. Gaunt.
\newblock The deferred approach to the limit.
\newblock \emph{Philosophical Transactions of the Royal Society of London.
  Series A, containing papers of a mathematical or physical character},
  226\penalty0 (636--646):\penalty0 299--361, 1927.

\bibitem[Roache(1994)]{roache94}
P.~J. Roache.
\newblock Perspective: a method for uniform reporting of grid refinement
  studies.
\newblock \emph{Journal of Fluids Engineering}, 116\penalty0 (3):\penalty0
  405--413, 1994.

\bibitem[Roache(1997)]{Roachdisc}
P.~J. Roache.
\newblock Quantification of uncertainty in computational fluid dynamics.
\newblock \emph{Annual Review of Fluid Mechanics}, 29\penalty0 (1):\penalty0
  123--160, 1997.

\bibitem[Roache(1998)]{Roache1}
P.~J. Roache.
\newblock Verification of codes and calculations.
\newblock \emph{AIAA Journal}, 36\penalty0 (5):\penalty0 696--702, 1998.

\bibitem[Sacks et~al.(1989{\natexlab{a}})Sacks, Schiller, and Welch]{sacks89a}
J.~Sacks, S.~B. Schiller, and W.~J. Welch.
\newblock Designs for computer experiments.
\newblock \emph{Technometrics}, 31\penalty0 (1):\penalty0 41--47,
  1989{\natexlab{a}}.

\bibitem[Sacks et~al.(1989{\natexlab{b}})Sacks, Welch, J., and Wynn]{sacks89b}
J.~Sacks, W.~J. Welch, Mitchell~T. J., and H.~P. Wynn.
\newblock Design and analysis of computer experiments.
\newblock \emph{Statistical Science}, 4\penalty0 (4):\penalty0 409--435,
  1989{\natexlab{b}}.

\bibitem[Stein(1999)]{stein99}
M.~L. Stein.
\newblock \emph{Interpolation of Spatial Data: Some Theory for Kriging}.
\newblock Springer Series in Statistics. Springer, New York, 1999.

\bibitem[Stroh et~al.()Stroh, Bect, Demeyer, Fischer, and
  Vazquez]{stroh2017AMCTM}
R.~Stroh, J.~Bect, S.~Demeyer, N.~Fischer, and E.~Vazquez.
\newblock \emph{Integrating hyper-parameter uncertainties in a multi-fidelity
  {B}ayesian model for the estimation of a probability of failure}, pages
  349--356.
\newblock World Scientific.

\bibitem[Stroh et~al.(2017)Stroh, Bect, Demeyer, Fischer, Marquis, and
  Vazquez]{stroh2017firesafety}
R.~Stroh, J.~Bect, S.~Demeyer, N.~Fischer, D.~Marquis, and E.~Vazquez.
\newblock Assessing fire safety using complex numerical models with a
  {B}ayesian multi-fidelity approach.
\newblock \emph{Fire Safety Journal}, 91:\penalty0 1016--1025, 2017.

\bibitem[Tuo et~al.(2014)Tuo, Wu, and Yu]{twy2014}
R.~Tuo, C.~F.~Jeff Wu, and D.~Yu.
\newblock Surrogate modeling of computer experiments with different mesh
  densities.
\newblock \emph{Technometrics}, 56\penalty0 (3):\penalty0 372--380, 2014.

\bibitem[Weaver et~al.(1990)Weaver, Timoshenko, and Young]{timo}
W.~Weaver, S.~P. Timoshenko, and D.~H. Young.
\newblock \emph{Vibration Problems in Engineering}.
\newblock John Wiley \& Sons, 1990.

\bibitem[Welch et~al.(1992)Welch, Buck, Sacks, Wynn, Mitchell, and
  Morris]{Welch92}
W.~J. Welch, R.~J. Buck, J.~Sacks, H.~P. Wynn, T.~J. Mitchell, and M.~D.
  Morris.
\newblock Screening, predicting and computer experiments.
\newblock \emph{Technometrics}, 34:\penalty0 15--25, 1992.

\end{thebibliography}
  
\end{document}